\documentclass[11pt]{iopart}
\usepackage{hyperref}
\usepackage{amssymb}
\usepackage{amsopn}

\newcommand{\Planck}{M_{\mathrm{P}}}

\renewcommand{\geq}{\geqslant}
\newcommand{\fnl}{f_{\mathrm{NL}}}
\newcommand{\cl}{\mathrm{cl}}
\newcommand{\intr}{\mathrm{int}}
\renewcommand{\etal}{\emph{et al.}}

\begin{document}
	\title{Relics of spatial curvature in the primordial non-gaussianity}
	\date{\today}
	\author{Tim Clunan and David Seery$^1$}
	\address{Department of Applied Mathematics and Theoretical Physics \\
	Centre for Mathematical Sciences \\
	Wilberforce Road, Cambridge, CB3 0WA, United Kingdom}
	\eads{\mailto{T.P.Clunan@damtp.cam.ac.uk} and
		\mailto{D.Seery@damtp.cam.ac.uk}}
	\pacs{98.80.-k, 98.80.Cq, 11.10.Hi}
	\begin{abstract}
		We study signatures in the Cosmic Microwave Background (CMB)
		induced by the presence of
		strong spatial curvature prior to the epoch of
		inflation which generated our present universe.
		If inflation does not last sufficiently long to drive the
		large-scale spatial curvature to zero, then presently observable
		scales may have left the horizon while spatial slices
		could not be approximated by a flat, Euclidean geometry.		
		We compute corrections to the power spectrum and
		non-gaussianity of the CMB temperature anisotropy
		in this scenario.
		The power spectrum does not receive significant corrections
		and is a
		weak diagnostic of the presence of curvature in 
		the initial conditions, unless its running can be determined
		with high accuracy.
		However, the bispectral non-gaussianity parameter
		$\fnl$ receives modifications on the largest
		observable scales.
		We estimate that the maximum signal would correspond to
		$\fnl \sim 0.3$, which is out of reach for present-day
		microwave background experiments.
	\vspace{3mm}
	\begin{flushleft}
		\textbf{Keywords}:
		Inflation,
		Cosmological perturbation theory,
		Physics of the early universe,
		Quantum theory in curved spacetime.
	\end{flushleft}
	\end{abstract}
	\maketitle

\section{Introduction}
\label{sec:introduction}
The precise data which led to our present era of observation-driven
cosmology have largely been taken from measurements of the
temperature and polarization anisotropies
of the Cosmic Microwave Background \cite{Dunkley:2008ie,Komatsu:2008hk}.
These anisotropies were apparently sourced by large-scale fluctuations
of primordial origin, which passed inside the causal horizon while the
universe was dominated by a hot plasma of relativistic matter
and radiation. The existence of these
long wavelength fluctuations can be inferred today
by studying the imprint of the plasma oscillations which they seeded
and which were left intact when radiation subsequently decoupled from
matter.

The origin of these superhorizon fluctuations is uncertain.
With present-day observations, their properties are indistinguishable
from a precisely gaussian random field.
It is possible they were synthesized during an early epoch of inflation,
in which case the detectability of non-gaussian features is known to be
a precise discriminant between the simplest model, characterized by a
single degree of freedom \cite{Maldacena:2002vr}
and more complicated models, characterized by many degrees of
freedom \cite{Lyth:2005fi}.
(For example, see
Refs.~\cite{Lyth:2005qk,Alabidi:2006wa,
Sasaki:2008uc,Naruko:2008sq,Byrnes:2008wi,Byrnes:2008zy}.)
In addition to its dependence on the dynamics of the inflationary phase,
the non-gaussian signal may
be sensitive to the composition of the universe
before and after inflation.
Subsequent to the end of the inflationary era, its evolution has been
explored by several authors
\cite{Bartolo:2003gh,Bartolo:2006cu,
Bartolo:2006fj,Pitrou:2008ak,Bartolo:2008sg,Boubekeur:2009uk}.
On the other hand, if inflation persisted for some number of e-folds
prior to observable scales leaving the horizon then the impact of
any pre-inflationary physics can be expected to be quite negligible
\cite{Wald:1983ky}.

If inflation did not last long enough to erase all memory of
earlier times, what information could be expected to survive in
the statistics of large-scale fluctuations?
Chen {\etal} \cite{Chen:2006xjb}
studied the impact of excitations above the Minkowski
vacuum%
	\footnote{When promoted to de Sitter space, the choice of boundary
	conditions which reproduces the Minkowski vacuum on small scales
	is usually known as the Bunch--Davies state
	\cite{Bunch:1978yq}.}
and found them to have a distinctive momentum dependence. Moreover,
under certain assumptions the traces of these excitations could be made
large. More detailed investigations were subsequently carried out
in Refs.~\cite{Holman:2007na,Meerburg:2009ys}.
Statistical features of this sort could probe
an enhancement of the correlations predicted by conventional inflationary
models due to perturbative effects,
but would necessarily have limited reach to detect
a coherent modification of the background spatial geometry.
In the conventional picture, where inflation is taken to have lasted
for many e-folds prior to horizon exit of observable scales,
the spatial geometry is taken to be flat to high precision.
If this assumption is relaxed, however, then the background may be
different. To study such a scenario one requires a new set of
propagators and vertices, which are the elements out of which
perturbation theory is constructed. Predictions obtained in this way could
quite easily be rather different to conventional results.
In the case of both perturbative or non-perturbative enhancements
the effect diminishes as spatial slices inflate and
isotropize, unless the modification is associated with
new physics at an energy somewhat higher than the
inflationary scale.%
	\footnote{If there is a sensitivity to new ultra-violet physics, then
	quantum excitations can have correlations which are not
	well-described by the standard formulation. Such modifications are
	not inflated away, but instead become imprinted in each
	new mode which crosses the horizon
	\cite{Chen:2006xjb,Holman:2007na,Meerburg:2009ys}.}

According to present ideas the pre-inflationary phase could have been
characterized by strong spatial disorder, of which the simplest manifestation
would be non-vanishing spatial curvature. Assuming
dark energy to be non-dynamical, observation presently
requires $-0.0178 < \Omega_k < 0.0066$ at $95\%$ confidence,
where $\Omega_k$ is evaluated today and the constraints are obtained
by combining the WMAP 5-year dataset with baryon acoustic oscillation
and supernova data \cite{Komatsu:2008hk}.
As observations improve the allowed range of $\Omega_k$ can be
expected to diminish, but if it remains consistent with zero
it will presumably be impossible to rule out a small positive
or negative value.
One might be led to seek more information in the statistical properties
of the large-scale perturbations, but
it is already known that curvature is hard to detect in the power spectrum
alone \cite{Halliwell:1984eu}.
For that purpose it seems worthwhile to explore whether the situation is
better for higher-order correlations.
Thus, both to decide what can be learned about the
pre-inflationary initial conditions,
and to determine whether curvature at the level
$|\Omega_k| \sim 10^{-2}$
can be probed by non-gaussian statistics,
one would like to obtain the inflationary bispectrum computed with
non-flat spatial slices.
In this paper we focus on the case of positive spatial curvature,
corresponding to $S^3$ spatial slices, since in that case the observational
limit is more favourable, as discussed above.

How large an effect should we expect?
The longest observable wavelength in the CMB contributes to its dipole,
although one must be wary in extracting data at low multipoles because
of scatter due to cosmic variance. Nevertheless, as a simple estimate,
a calculation of the comoving distance to the CMB dipole (or $\ell=2$ mode;
cf. Ref.~\cite{Freivogel:2005vv}) gives 
the diameter of the last scattering sphere as
\begin{equation}
	\fl
	d_{\ell=2}=2\int_{t_{\mathrm{dc}}}^{t_{0}}
	\frac{d t}{e^{\rho}} = 2
	\int_{(1+z_{\mathrm{dc}})^{-1}}^{1}
	\frac{d x}{x \sqrt{-1 + x^2 (\Omega_{m}
	x^{-3}+\Omega_{\Lambda}+\Omega_{r} x^{-4})/|\Omega_k|}} ,
\end{equation}
where we have specialized to $S^3$ slices and, as above,
the $\Omega_{i}$ are evaluated today (at
time $t_0$).
The decoupling time is denoted $t_{\mathrm{dc}}$ and
$e^{\rho}$ is the scale factor.
For the purposes of making an estimate, let us assume that the present-day
spatial curvature takes the WMAP5+BAO+SN central value
(at $68\%$ confidence)
$\Omega_k \approx -0.0050$ \cite{Komatsu:2008hk}.
The remaining cosmological parameters can be chosen equal to the
best-fit concordance model where
$\Omega_\Lambda = 0.721$,
$\Omega_r = 8 \times 10^{-5}$
and the CMB decouples at a redshift $z_{\mathrm{dc}} = 1100$.
This gives $d_{\ell=2} \approx 0.46$. On the other hand, in these units,
the radius of curvature
of the spatial slices is set by the scale $r = 1$.
It follows that the CMB dipole receives contributions from wavelengths
which are a significant fraction of the radius of curvature.
Where this is the case, we do not expect
to be able to neglect spatial curvature in calculating
microwave background observables.

In this paper we estimate the observational signatures
of this scenario. We study the correction to the power spectrum
from single-field, slow-roll inflation
assuming that inflation began while $\Omega_k$ was non-negligible,
and then consider the implications for the
three-point function of the curvature perturbation. 
Our result agrees with the known flat-slicing result
\cite{Maldacena:2002vr} in the small-scale limit $k \gg 1$
(where $k$ is the wavenumber of
the mode under consideration), in agreement with
the na\"{\i}ve expectation that curvature should play a less important
role as inflation proceeds. On the other hand for small $k$
the fluctuations can feel the curvature of the $S^3$ slices, leading
to terms which are larger than the flat $\mathbb{R}^3$
prediction by a factor
of $1/\epsilon$, where $\epsilon \equiv \dot\phi^2 / 2 \dot\rho^2$
is the slow-roll parameter which measures the deviation of the universe
from exact de Sitter space.
This enhancement is driven by quantum interference among the
fluctuations at horizon-crossing, and is maximized close to the
equilateral configuration where all momenta participating in the
three-point function have an approximately equal magnitude.%
	\footnote{Therefore,
	we do not expect that this enhancement by $1/\epsilon$
	would provide an explanation for the observation of large
	non-gaussianity reported by Yadav \& Wandelt \cite{Yadav:2007yy}.}

The layout of this paper is as follows. In \S\ref{sec:action} we
compute the action for a scalar field coupled to Einstein gravity
in de Sitter space with $S^3$ spatial hypersurfaces. After including the
effect of small perturbations in the metric, this controls how
small fluctuations interact and become non-linear as
the expansion of the universe draws them across the causal horizon.
In order to obtain a description of the non-linearities which are
observable in the CMB bispectrum,
it is sufficient to expand the action to third order in the perturbations.
In \S\ref{sec:power} we study corrections to the two-point function, and
show that curvature plays essentially
no role in determining its structure at tree level.
In \S\ref{sec:third_order_action} we give the third order action and
in \S\ref{sec:three-point} we calculate the three point function.
We then discuss the non-gaussian parameter
$\fnl$ imprinted in this scenario
and show that we have the expected limit for modes of
small wavelength.
Finally, we give our conclusions in \S\ref{sec:conclusions}.

Throughout this paper,
we work in units where the $\hbar = c = 1$ and the reduced Planck
mass is normalized to unity, giving $\Planck^{-2} \equiv 8 \pi G = 1$,
where $G$ is Newton's gravitational constant.
The unperturbed background metric is
\begin{equation}
	d s^2 = - d t^2 + e^{2 \rho} \gamma_{ij} \, d x^i \, d x^j ,
\end{equation}
and $\gamma_{ij}$ is the metric on the unit three-sphere.
This choice means that the constant curvature $k$ is normalized to unity,
and to avoid clutter we will omit writing $k$ explicitly.

\section{Computation of the action}
\label{sec:action}

To study the evolution of small fluctuations around the time of
horizon exit, one must couple gravity to
whichever fields are responsible for driving inflation.
Since the energy density in these fields
dominates the energy budget of the universe by assumption, small
fluctuations necessarily induce perturbations in the background metric.
The non-linearity of gravity endows these perturbations with
self-interactions which cannot be neglected.

To simplify the subsequent calculation, it is very convenient
to take the perturbed metric in the Arnowitt--Deser--Misner or
so-called \emph{ADM} form \cite{Arnowitt:1960es}, where
\begin{equation}
d s^{2}=-N^{2}d t^2 + h_{i j} (d x^{i}+N^{i} d t) (d x^{j}+N^{j} d t) .
\end{equation}
In this equation $N$ and $N^i$ are the lapse function and shift-vector,
whose function is to guarantee reparametrization invariance of
the metric. They are
not propagating degrees of freedom but are instead removed by
constraint equations coming from the Einstein action;
and $h_{ij}$ is the perturbed metric on spatial slices,
\begin{equation}
	h_{ij} = e^{2(\rho + \zeta)} ( \gamma_{ij} + \Gamma_{ij} ) .
	\label{eq:gauge-choice}
\end{equation}
The quantity $\zeta$ is known as the \emph{curvature perturbation}.
If gravitational waves are present, represented here by a
transverse and traceless first-order tensor $\Gamma_{ij}$,
then they should be included
as corrections to the unit three-sphere metric $\gamma_{ij}$.
We have implicitly adopted a spatial gauge in order to write
Eq.~\eref{eq:gauge-choice}.
In the present paper we focus solely on the properties of scalar
degrees of freedom, which can be summarized in terms of the
curvature perturbation, and set $\Gamma_{ij} = 0$.
This is adequate for the purpose of computing the bispectrum of $\zeta$.
In the absence of perturbations,
the lapse function and shift-vector
reduce to unity and zero, respectively,
and otherwise can be expressed algebraically in terms of the propagating
degrees of freedom.

The action is taken to be the standard Einstein--Hilbert term,
together with a real scalar field $\phi$, supplemented with
the York--Gibbons--Hawking boundary term,
\begin{equation}
	\fl\nonumber
	S = \frac{1}{2} \int d t \, d^3 x \; \sqrt{h} \Bigg\{
		N \left( R^{(3)} - 2 V - h^{ij} \partial_i \phi \partial_j \phi
		\right) \\ \mbox{} + \frac{1}{N} \left( E_{ij} E^{ij} - E^2 +
			[ \dot{\phi} - N^i \partial_i \phi ]^2
		\right) \Bigg\} ,
	\label{eq:action}
\end{equation}
where $E_{ij}$ is a rescaled version of the extrinsic curvature,
defined by $K_{ij} \equiv E_{ij}/N$ and given explicitly by
\begin{equation}
	E_{i j}
	\equiv
	\frac{1}{2} (\dot{h}_{i j}-\nabla_{i} N_{j}-\nabla_{j} N_{i}) .
\end{equation}
The scalar $E$ is its trace, obeying $E \equiv E^i_i$, where spatial
indices are raised and lowered with the metric $h_{ij}$.
The covariant derivative compatible with
this metric
is denoted $\nabla_{i}$ and an overdot denotes a derivative with
respect to cosmic time, so that $\dot{x} \equiv d x / d t$.
There are two scalar degrees of freedom in this system: the
fluctuations $\zeta$ and $\delta \phi$.
However, only one of these is a propagating mode; the other is a
gauge degree of freedom and can be removed by a coordinate redefinition.
Although a wide variety of gauge choices exist \cite{Kodama:1985bj}, two
are of special interest. The first is \textit{comoving gauge}, where
the scalar perturbation is set to zero and only the curvature
perturbation propagates,
\begin{equation}
	\label{comoving_gauge}
	\delta \phi = 0, \quad \mbox{and} \quad
	h_{i j}=e^{2\rho +2\zeta}\gamma_{i j} .
\end{equation}
Late on when a mode has passed outside the horizon $\gamma_{ij}$ can be 
effectively replaced by a flat metric on $\mathbb{R}^3$ since the
equations of motion are negligibly affected by
the curvature of the $S^3$ slices.
In this limit 
the scale of the metric is arbitrary and the
equations of motion possess a symmetry under 
dilatations of the scale factor, where $\rho \mapsto \rho + \alpha$
for any constant $\alpha$. It follows that on very large scales
$\zeta$ must be effectively conserved,
because it becomes indistinguishable from a shift in the
background value of $\rho$. This conservation law has been shown to
follow---independently of the theory of gravity in
question \cite{Lyth:2004gb}---from energy conservation and the absence of
non-adiatbatic pressure.
It is therefore convenient to express the answer in
terms of $\zeta$ when outside the horizon.

Unfortunately, the calculation of correlation functions directly in
comoving gauge is technically complicated. It is more convenient to work
in the \textit{uniform curvature gauge}, which is defined by
\begin{equation}
	\label{deltaphi}
	\delta\phi = \varphi, \quad \mbox{and} \quad
	h_{i j}=e^{2 \rho}\gamma_{i j}.
\end{equation}
In this gauge it is simple to compute the correlation functions of
$\varphi$, but more work is needed to
make a gauge transformation into the
observationally-relevant variable $\zeta$. It is this approach which we will
follow in the remainder of this paper.

\subsection{Constraint equations}

The action~\eref{eq:action} does not depend on time derivatives of $N$
or $N^i$, and therefore its variation with respect to these fields does
not produce evolution equations but only constraints.
We need to solve these constraint equations to first order%
	\footnote{As noted in
	Ref.~\cite{Maldacena:2002vr} and given in more detail in
	Ref.~\cite{Chen:2006nt}, the higher order terms in $N$
	and $N^{i}$ occur in the action multiplying the
	Hamiltonian and momentum constraints respectively,
	and hence they can be neglected.}.
The Hamiltonian constraint is
\begin{equation}
	R^{(3)} - 2V - \frac{1}{N^2} (E_{i j} E^{i j} -E^{2})
	- \frac{1}{N^2} (\dot{\phi}-N^{i} \phi_{,i})^{2}
	- h^{i j} \phi_{,i} \phi_{,j} = 0 ,
\end{equation}
where we have adopted the convention that a partial derivative with
respect to $x^i$ is denoted with a comma, so that $\phi_{,j} \equiv
\partial_j \phi$.
The momentum constraint is
\begin{equation}
	\nabla_{i} \left\{ \frac{1}{N} (E_{j}^{i}-\delta_{j}^{i} E) \right\}
	- \frac{1}{N} (\dot{\phi} -N^{i} \phi_{,i}) \phi_{,j} = 0
\end{equation}
In subsequent expressions, to simplify the notation, we will use
$\phi=\phi (t)$ to denote the background value of the field
and reserve $\varphi$ for its perturbation.

One can now
specialize to the uniform curvature gauge.
In this gauge, the solution to the constraints takes the 
form $N \equiv 1 + \delta N$, where $\delta N$ obeys
\begin{equation}
	\delta N \equiv \frac{1}{\dot{\rho}}
	\Big( \frac{\dot{\phi}}{2} \varphi + \chi \Big) ;
\end{equation}
and $N_{i} \equiv \chi_{,i}$, where $\chi$ is an auxiliary quantity
determined by
\begin{equation}
	\label{chi}
	\chi \equiv \epsilon (\triangle+3-\epsilon)^{-1} \Big[\frac{d}{d
	t}\Big(-\frac{\dot{\rho}}{\dot{\phi}}\varphi
	\Big)+\frac{e^{-2\rho}}{\dot{\phi}}\varphi\Big]
\end{equation}
and $\triangle$ is the laplacian on the unit three-sphere.
We note that both $\delta N$ and $N^i$ are suppressed by the square
root of a slow-roll parameter,
	\begin{equation}
	\label{order}
	\delta N = \Or (\sqrt{\epsilon}) \quad \mbox{and} \quad
	N^{i} = \Or (\sqrt{\epsilon}) ,
\end{equation}
where $\epsilon \equiv \dot{\phi}^2/2 \dot{\rho}^2$ is the
slow-roll parameter defined in \S\ref{sec:introduction}.
This scaling will be extremely important
when we come to extract the terms of lowest order in the
slow-roll approximation from the action.

\subsection{The second order action}

In the uniform curvature gauge, Eq.~\eref{deltaphi},
the second order term in the action is
\begin{eqnarray}
	\fl\nonumber
	S_{2}=\frac{1}{2}\int d t \, d^{3} x \; e^{3\rho}\sqrt{\gamma} \;
	\Big[-V_{0}''\varphi^{2}-\frac{\dot{\phi}}{\dot{\rho}} V_{0}'
	\varphi^{2}-\frac{2}{\dot{\rho}} V_{0}' \varphi \chi +
	\dot{\varphi}^{2}-e^{-2\rho} \gamma^{i j}\varphi_{,i}\varphi_{,j}\\
	\mbox{}-2\chi 
	\triangle \chi - \frac{\dot{\phi}^{2}}{\dot{\rho}}\varphi
	\dot{\varphi}-\frac{2\dot{\phi}}{\dot{\rho}}\chi\dot{\varphi}+\Big(
	\frac{\dot{\phi}^{2}}{\dot{\rho}^{2}}-6 \Big)
	\Big(\frac{\dot{\phi}^{2}}{4}\varphi^{2}+\dot{\phi}\chi\varphi+\chi^{2}
	 \Big)\Big]
\end{eqnarray}
where $\chi$ is given by Eq.~\eref{chi}
and $\triangle$ is again the laplacian on the unit three-sphere. Since $V_{0}'' =-3 e^{-2\rho}+O(\epsilon)$ and \eref{order} holds we have at 
lowest order in slow roll
\begin{equation}
	S_{2}=\frac{1}{2}\int d t \, d^{3} x \; e^{3\rho}\sqrt{\gamma} \;
	[\dot{\varphi}^{2}+3 e^{-2\rho}\varphi^{2}-e^{-2\rho}\gamma^{ij}
	\varphi_{,i}\varphi_{,j}] .
\end{equation}
The field equation for $\varphi$ which follows from this action is
\begin{equation}
	\label{initialeom}
	\frac{d}{d t} \Big( e^{3 \rho}\frac{d}{d t}\varphi \Big) -
	3 e^{\rho}\varphi - e^{\rho}\triangle \varphi = 0.
\end{equation}
As discussed below Eq.~\eref{comoving_gauge},
the terms depending on curvature scale out of Eq.~\eref{initialeom}
in the limit $e^{\rho} \rightarrow \infty$ after which
Eq.~\eref{initialeom} has a time-independent solution.

\section{Corrections to the power spectrum from curvature}
\label{sec:power}
\subsection{Scalar mode functions on the three-sphere}

We wish to solve Eq.~\eref{initialeom} at lowest order in the slow-roll
approximation. For this purpose it is first necessary
to obtain an expression for $e^{\rho}$ at
lowest order. One does this by Taylor expanding the scale factor about a particular time and collecting the terms of lowest order.
In principle, the Taylor expansion can be performed at any point.
However, because
our final answer will involve an integral which receives its 
largest contribution from around the time of horizon crossing,
we can achieve the greatest accuracy by expanding around the
time of horizon exit. This is precisely analogous to the expansion of
$H$ and other background quantities around the time of horizon exit in
the standard calculation on flat spatial slices.

The background equations for the scale factor
consist of the Friedmann constraint,
\begin{equation}
	3 \dot{\rho}^{2} = 
	\frac{1}{2} \dot{\phi}^{2}+V_{0}-3 e^{-2\rho} ,
	\label{eq:friedmann}
\end{equation}
together with the Raychaudhuri equation
\begin{equation}
	\ddot{\rho}=
	- \frac{1}{2}\dot{\phi}^{2}+e^{-2\rho} .
\end{equation}
In addition,
the scalar field evolves according to the conventional Klein--Gordon
equation
\begin{equation}
	\ddot{\phi}+3 \dot{\rho}\dot{\phi}+V_{0}' = 0 ,
\end{equation}
which is not corrected in the presence of curvature. At lowest order in slow roll we can write the $n$th order
time derivative of the scale factor in terms of alternating expressions
for odd and even $n$,
\begin{eqnarray}
	\left.\left( \frac{d}{d t} \right)^{2 n}e^{\rho}
	 \right|_{t=t_{*}}=\,e^{\rho_{*}} (\dot{\rho}_{*}^{2}+e^{-2 
	\rho_{*}})^{n} + \Or(\epsilon) ; \\
	\left.\left( \frac{d}{d t} \right)^{2 n+1}e^{\rho} 
	\right|_{t=t_{*}}=\,e^{\rho_{*}} \dot{\rho}_{*} (\dot{\rho}_{*}^{2}+
	e^{-2 \rho_{*}})^{n}+\Or(\epsilon) .
\end{eqnarray}
where an asterisk $\ast$ denotes evaluation at the time of horizon
crossing for the corresponding wavenumber $k$.
It follows that after expansion around the time
$t = t_\ast$, the scale factor can be
expressed at lowest order in the slow-roll approximation by
\begin{equation}
	\fl
	e^{\rho} \simeq e^{\rho_{*}}\cosh [(t-t_{*})\sqrt{\dot{\rho}_{*}^{2}+
	e^{-2 \rho_{*}}}]
	+ \frac{\dot{\rho}_{*} e^{\rho_{*}}}
		{\sqrt{\dot{\rho}_{*}^{2}+e^{-2 \rho_{*}}}}
	\sinh [(t-t_{*})\sqrt{\dot{\rho}_{*}^{2}+e^{-2 \rho_{*}}}] .
	\label{eq:approx-scale}
\end{equation}
We can see that this is the scale factor for a de Sitter spacetime with
$\rho$ and $\dot{\rho}$ matched to our slowly rolling background
at $t=t_{*}$. In fact, it is usually more convenient to work in terms
of a conformal time variable, $\eta$, defined by
$\eta = \int_{t_{\mathrm{throat}}}^t d t' / e^\rho(t')$,
where $t_{\mathrm{throat}}$ corresponds to the throat of the de Sitter
hyperboloid.
In terms of the $\eta$ variable,
Eq.~\eref{eq:approx-scale} can be written
\begin{equation}
	\label{conf_sc_fact}
	e^{\rho}=
	\frac{1}{\dot{\rho}_{*} \lambda \cos \eta}+\Or(\epsilon) .
\end{equation}
It follows that
in the neighbourhood of $\eta = \eta_\ast$ the Hubble rate obeys
\begin{equation}
	\dot\rho = \dot\rho_* \lambda \sin \eta+\Or(\epsilon),
\end{equation}
where $\lambda\equiv\sqrt{1+e^{-2 \rho_{*}} / \dot{\rho}_{*}^{2}}$
and $\eta \in (-\pi / 2, \pi / 2 ) $ for $t\in \mathbb{R}$.
The relation between conformal and physical time is 
\begin{equation}
	\tan \frac{\eta}{2} = \tanh \frac{1}{2} \Big[(t-t_{*})\dot{\rho}_{*}
	\lambda + \tanh^{-1} \Big(\frac{1}{\lambda}\Big)\Big] 
\end{equation}
Combining Eqs.~\eref{initialeom} and~\eref{conf_sc_fact},
we find that the equation for the modes at lowest order in slow roll is 
\begin{equation}
	\label{eom}
	\varphi_{k}'' +2 \tan \eta\; \varphi_{k}' + (k^2 -4) \varphi_{k} = 0
\end{equation}
where a prime $'$ denotes a derivative with respect to $\eta$,
and $-(k^2 -1)$, with $k \in \mathbb{N} - \{0\}$, are the eigenvalues of the
laplacian $\triangle$.  This is solved by the 
normalised positive mode \cite{Birrell:1982ix}
constructed to vanish at the south pole of the Hawking--Moss instanton
\cite{Hawking:1982my}
describing de Sitter space at horizon crossing,
\begin{eqnarray}
	\fl\nonumber
	\varphi_{k}^{\cl}= \,
	\frac{\dot{\rho}_{*} \lambda (k^{2}-3)^{1/4}}
	{\sqrt{2 (k^{2}-4)}} \Bigg[ F \Big(
	\begin{array}{cc}
		- 1/2 + \sqrt{k^2-3}/2; & - 1/2 - \sqrt{k^2-3}/2 \\
		\multicolumn{2}{c}{1/2}
	\end{array}
	\Big| \sin^{2}\eta \Big) \\
	\label{mode}
	\mbox{} + \imath \,
	\frac{(k^2-4)}{\sqrt{k^2-3}}\sin \eta \, F \Big(
	\begin{array}{cc}
		\sqrt{k^2-3}/2; & -\sqrt{k^2-3}/2 \\
		\multicolumn{2}{c}{3/2}
	\end{array}
	\Big| \sin^{2}\eta \Big)\Bigg]
\end{eqnarray}
where $F$ is the Gauss
hypergeometric function and $\imath^2 \equiv -1$.
The south pole is at $\eta= \imath \infty$.
This is equivalent to the usual formulation in which the positive mode
is taken to approach its Minkowski value deep inside the horizon, where
it is insensitive to the curvature of spacetime.

Ultimately we wish to use these mode functions to obtain analytic
estimates of the
$n$-point expectation values of fluctuations in $\varphi$
around the time of horizon crossing. However,
the presence of hypergeometric functions in Eq.~\eref{mode}
means that it is awkward to use such expressions for this purpose.
To obtain a more useful approximation, one may first re-write the equation of motion, Eq.~\eref{eom}, as
\begin{equation}
	\left(\frac{\varphi_{k}}{\cos\eta}\right)'' = - (k^2 -5 -2 \tan^2 \eta )
	\frac{\varphi_{k}}{\cos\eta} .
\end{equation}
The equation admits an ``almost-WKB'' solution, corresponding to
\begin{equation}
	\label{almost_wkb}
	\varphi_{k} \approx \frac{\dot{\rho}_{*} \lambda}{\sqrt{2 k}}
	e^{\imath k \eta} \cos \eta ,
\end{equation}
which becomes a good approximation for an increasingly large neighbourhood
of $\eta = 0$ as $k \rightarrow \infty$. Unfortunately, we cannot
conclude that Eq.~\eref{almost_wkb} is sufficient to obtain an
acceptable estimate of the three-point expectation value of scalar
fluctuations near horizon crossing (where $\eta \rightarrow \pi/2$),
because it freezes out to the wrong asymptotic value.
This is a fatal deficiency. It would lead to an unreliable
estimate of the magnitude of the bispectrum and therefore untrustworthy
conclusions regarding the observational relevance of $\fnl$.
To find an approximate form for the mode which gives a good approximation
over the \emph{entire} relevant range of $\eta$,
we can use standard results relating to hypergeometric functions to find that at late times $\eta \rightarrow \pi/2$ the field freezes out to 
\begin{equation}
	\varphi_{k}^{\cl} \rightarrow
	- \frac{\imath \dot\rho_* \lambda}
	{(k^2-3)^{1/4} \sqrt{2 (k^2-4)}}
	e^{\imath \pi \sqrt{k^2-3} / 2}
	\approx
	- \frac{\imath \dot \rho_* \lambda}{\sqrt{2 k^3}}
	e^{\imath \pi k/2},
\end{equation}
where the approximation holds for large $k$.
It follows that if we take $\varphi_k^{\cl}$ to be given by
\begin{equation}
	\label{mode_approx}
	\varphi_{k}^{\cl} \approx
	\frac{\dot{\rho}_{*} \lambda}{\sqrt{2 k}}
	\Big(\cos \eta - \frac{\imath}{k} \Big) e^{\imath k \eta} ,
\end{equation}
then we obtain the correct value and derivative at $\eta=\pi / 2$,
and indeed this gives a good estimate for all $\eta$.

In Eq.~\eref{mode_approx} and below, the label
``$\cl$'' indicates that this set of mode functions form a good
basis out of which we can build a quantum field when we come to quantize
this theory. To do so, one introduces creation
and annihilation operators $a^\dag_{\vec{k}}$ and $a_{\vec{k}}$
which respectively create and destroy particles
(as measured by an inertial observer deep inside the horizon)
with momentum $\vec{k}$. 
The quantum field corresponding to $\varphi$ can be constructed
by the usual canonical procedure, leading to
\begin{equation}
	\label{field1}
	\varphi(\eta, \vec{x}) = \sum_{\vec{k}} \Big( Q_{\vec{k}}(\vec{x})
	\,\varphi_{k}^{\cl}(\eta) a_{\vec{k}}^{\dagger} + Q_{\vec{k}}^{*}(\vec{x})
	\,\varphi_{k}^{\cl *}(\eta) a_{\vec{k}}\Big)
\end{equation}
where the $S^3$ harmonics $Q_{\vec{k}}$ are defined
by $\triangle Q_{\vec{k}}=-(k^2 -1) Q_{\vec{k}}$.
The case $k=1$ is the homogeneous background and the case $k=2$ gives
purely gauge modes.%
	\footnote{To see that the $k=2$ modes are pure gauge,
	suppose $\varphi_{2}$ satisfies
	$\triangle \varphi_{2} =-3 \varphi_{2}$. It must also satisfy
	$\varphi_{2 | i j}=-\gamma_{i j} \varphi_{2}$,
	as can be verified by checking each $k=2$ mode,
	and therefore
	it is possible to perform a gauge transformation from 
	the metric
	\begin{equation}
		d s^2 = -N^2 d t^2 + e^{2 \rho} \gamma_{i j}
		(d x^i +N^i d t) (d x^j +N^j d t), \quad \varphi=\varphi_{2}
	\end{equation}
	to a superficially different but gauge-equivalent metric
	\begin{equation}
		d s^2 = -\tilde N^2 d \tilde t^2 + e^{2 \rho} \gamma_{i j}
		(d \tilde x^i +\tilde N^i d \tilde t)
		(d \tilde x^j +\tilde N^j d \tilde t), \quad
		\tilde \varphi \equiv 0 .
	\end{equation}
	The coordinate transformation which allows us to pass between these
	two equivalent forms can be written
	\begin{equation}
		\tilde t = t + \frac{\varphi_2}{\dot{\phi}}, \quad
		\tilde x^i = x^i + \frac{\dot \rho}{2 \dot \phi} \varphi_{2}^{|i} .
	\end{equation}
	}
For the purposes of studying fluctuations neither of these modes are
relevant, and we will therefore always be
interested in modes of wavenumber three or larger, so that
$k \geq 3$.

\subsection{Appropriate coordinates}

We use coordinates in which the metric on $S^3$ takes the form
\begin{equation}
	d s_3^2 = d\chi^2+\sin^2 \chi \; d \Omega_2^2
\end{equation}
where $d \Omega_2^2$ is the usual metric on $S^2$. These coordinates are
convenient because if we take our position to be given by $\chi = 0$ then
the CMB can be thought of as located on the copy of
$S^2$ at $\chi=\chi_L$, where the subscript $L$ denotes
the time of last scattering.
In this form the harmonics $Q_{\vec{k}}$ are given by
\begin{equation}
	\label{harmonic_breakdown}
	Q_{\vec{k}} \equiv \Pi_{k}^{l} (\chi) Y_{l}^{m} (\theta , \phi) ,
\end{equation}
where $\vec{k} \equiv (k,l,m)$, the
$Y_{l}^{m}$ are the usual normalised harmonics on $S^2$ and 
	\begin{equation}
	\label{harmonic_breakdown_2}
	\Pi_k^l (\chi) = \sqrt{\frac{2 (k-1-l)! k}{\pi (k+l)!}}
	(2^l l! \sin^l \chi) C_{k-1-l}^{l+1} (\cos \chi) .
\end{equation}
In Eq.~\eref{harmonic_breakdown_2},
$C_{k-1-l}^{l+1}$ is a Gegenbauer polynomial and
$k-1 \geq l \geq | m |$. The list $\vec{k} = (k,l,m)$ labels the
quantum numbers associated with each harmonic, with $k$ labelling
the ``radial'' $\chi$-harmonic and $(l,m)$ the usual harmonic labels
on $S^2$ associated with the spherical harmonics $Y_{l}^m$. 
Under the exchange $m \mapsto -m$, the harmonics $Q_{\vec{k}}$ 
have the property%
	\footnote{The $(-1)^m$ term here comes from
	$[Y_l^m (\theta,\phi)]^* =(-1)^m Y_l^{-m} (\theta,\phi)$.}
that $Q_{\vec{k}}^* = (-1)^m Q_{-\vec{k}}$.
Therefore we may write the field (\ref{field1}) in the convenient form
\begin{equation}
	\varphi(\eta, \vec{x}) = \sum_{\vec{k}} Q_{\vec{k}}(\vec{x}) \Big[
	\,\varphi_{k}^{\cl}(\eta) a_{\vec{k}}^{\dagger} + (-1)^m
	\varphi_{k}^{\cl *}(\eta) a_{-\vec{k}} \Big]
\end{equation}

\subsection{The two-point function}
From the above we arrive at the two-point function evaluated at late times 
\begin{equation}
	\fl
	\langle \varphi_{\vec{k}_1}\varphi_{\vec{k}_2} \rangle = (-1)^{m_1}
	\delta_{-\vec{k}_1, \vec{k}_2} |
	\varphi_{k_1}^{\cl} (\pi /2) |^2 = (-1)^{m_1}
	\delta_{-\vec{k}_1, \vec{k}_2}
	\frac{\dot\rho_*^2 \lambda^2}{2 \sqrt{k_1^2 -3} (k_1^2 - 4)} .
\end{equation}
In order to sum expectation values of $\varphi$ into
expectation values of the curvature perturbation, $\zeta$, one can use
the non-linear $\delta N$ formalism, which will be described in more
detail in \S\ref{sec:deltaN} below. Here we merely quote the result for the
two-point function in order to demonstrate that the power spectrum
is consistent with the case of flat spatial slices, giving
\begin{equation}
	\label{two_pt}
	\langle \zeta_{\vec{k}_1}\zeta_{\vec{k}_2} \rangle
	= (-1)^{m_1} \delta_{-\vec{k}_1, \vec{k}_2}
	\frac{\dot\rho_*^4 }{2 \dot\phi_*^2 k_1^3} 
	\left[1+ O(k_1^{-2})\right]  \, .
\end{equation}
It follows that the power
spectrum of fluctuations in $\zeta$ is
the same as in the $\mathbb{R}^3$ case \cite{Halliwell:1984eu}, up to
very small corrections.

\subsection{The scalar spectral tilt}
In the absence of non-gaussianity, which will be studied in detail in
\S\S\ref{sec:third_order_action}--\ref{sec:three-point},
the principal discriminant among models
of inflation is the spectral tilt. Although Eq.~\eref{two_pt}
shows that the \emph{magnitude} of the power spectrum of fluctuations
approximately matches the flat space spectrum if $k$ is not small, the
tilt contains
curvature terms from $\lambda$ and $\ddot{\rho}_\ast$. These
curvature terms must be taken into account when determining whether the
model is compatible with the WMAP constraints on the spectral index $n_s$.

In the case of flat equal time slices one removes a term $1/k^3$ from the
two-point function when determining the tilt.
This is because the tilt is a measure of the deviation from 
scale invariance. Since the degeneracy of modes on $S^3$ with a given $k$
increases discretely with $k$, the notion of scale invariance is not natural on $S^3$;
however, experiment proceeds on the supposition of flat equal time slices and
thereby demands that we compute the tilt in the usual way.

For large $k$ the term $\dot\rho_*^{-2} e^{-2 \rho_*}$ is small and so the
tilt which follows from Eq.~\eref{two_pt} is
\begin{equation}
	\label{tilt}
	n_s - 1
	= 2 \eta_* -6 \epsilon_* +O(k^{-2})  \,   .
	\label{eq:large-k-tilt}
\end{equation}
where the slow-roll parameters $\epsilon$ and $\eta$ are defined by%
	\footnote{In the case of flat spatial slices, one often uses
	$\epsilon_H \equiv - \ddot\rho / \dot\rho^2$ 
	instead of the quantity $\epsilon$; here these differ by a curvature term,
	so that
	$\epsilon = \epsilon_H + \dot\rho^{-2} e^{- 2 \rho}$.
	The difference may be
	significant 
	for those modes passing outside the horizon when there is substantial
	curvature, so we are
	taking Eq.~\eref{eq:sr-def} to be the fundamental choice because the field
	is rolling slowly on a Hubble timescale. This also leads to 
	$(V')^2/2V^2\approx \epsilon / \lambda^4$ and $V''/V \approx \eta$.
	Note that $\epsilon_H$ need not be small 
	for modes passing out of the horizon early on.}
\begin{equation}
	\epsilon \equiv
	\frac{\dot\phi^2}{2\dot\rho^2}
	\quad \mbox{and} \quad
	\eta \equiv -
	\frac{\ddot\phi}{\dot\phi \dot\rho} +
	\frac{\dot\phi^2}{2 \dot\rho^2} .
	\label{eq:sr-def}
\end{equation}
The last term in Eq.~\eref{eq:large-k-tilt}
will be significantly smaller than the first 
two for $k\gtrsim 20$, and so we will recover
the usual result for flat spatial slices.
In this case it is known observationally that the tilt is small
\cite{Komatsu:2008hk}, so it follows that the
combination of $\epsilon_*$ and $\eta_*$ in Eq.~\eref{eq:large-k-tilt} is
too: generically, they are both small.
More generally, Eq.~\eref{eq:large-k-tilt} shows that the two-point function
is unlikely to be strongly sensitive to curvature unless good constraints
can be obtained on a possible running.

\subsection{Projecting onto the sky}
\label{sec:sky}
In order to properly compare results in the $S^3$ case with those in the
$\mathbb{R}^3$ case one should project onto the sky and compare the
resulting \emph{angular} expectation values of $\zeta$.
For this we need to consider the $S^3$ harmonics evaluated at $\chi_L$ on
the sky, where $\chi_L \ll 1$. From Eq.~\eref{harmonic_breakdown}
and the eigenvalue equation $\triangle Q_{\vec{k}} = -(k^2 - 1) Q_{\vec{k}}$
we see that $\Pi_k^l (\chi)$ obeys the defining equation
\begin{equation}
	\frac{1}{\sin^2 \chi}
	\frac{d}{d \chi} \Big( \sin^2 \chi \frac{d}{d \chi} \Pi_k^l(\chi \Big)
	- \frac{l(l+1)}{\sin^2 \chi} \Pi_k^l(\chi) + (k^2 -1)\Pi_k^l (\chi) = 0 .
\end{equation}
In the limit $\chi \ll 1$, $k \gg 1$ and $l \gg 1$ the solutions become
proportional to a spherical Bessel function, $\Pi_k^l(\chi) \propto
j_l(k\chi)$. The correct normalization can be obtained by studying
Eq.~\eref{harmonic_breakdown_2} in the
limit $\chi \rightarrow 0$ and yields
\begin{equation}
	\label{approx_pi}
	\Pi_k^l (\chi) \approx \sqrt{\frac{2}{\pi}} \,k \, j_l (k \chi)
\end{equation}
for $\chi \ll 1$, $k \gg 1$ and $l \gg 1$.
In the limit $\chi_L \ll 1$ the surface of last scattering is becoming
close to our point of observation, compared with the radius of curvature
of the $S^3$. We therefore expect that curvature ceases to play any role
and the two-point function projected onto the sky goes over to its flat
space counterpart. Indeed, one can show that in this limit (and with the
restriction $l \gg 1$ which guarantees we are looking on small angular
scales)
\begin{equation}
	\langle \zeta_{l_1 m_1} \zeta_{l_2 m_2} \rangle 
	\approx (-1)^{m_1} \delta_{l_1, l_2} \delta_{m_1, -m_2}
	\int_{0}^{\infty} \frac{d k}{k} \; \frac{\dot\rho_*^4}{\dot\phi_*^2}
	\frac{1}{\pi} \Big[ j_l (k \chi_L) \Big]^2
	\label{two-point-sky}
\end{equation}
where the approximation of continuing the integral to $0$ is valid
because the main contribution comes from the region $k\approx l \chi_L^{-1}$.
Eq.~\eref{two-point-sky} is equivalent to the well-known result in the
$\mathbb{R}^3$ case, which demonstrates the consistency of our calculation.

\section{The third-order action}
\label{sec:third_order_action}

In the previous section we expanded the action to second order in the
small fluctuation $\varphi$, which is sufficient to determine its
two-point statistics and therefore the power spectrum.
If we wish to
go further, however, and determine the leading non-linearity then it is
necessary to obtain a description of the process by which three
$\varphi$ quanta can interact. This information is provided by the third
order term in the expansion of the action, which we will determine in the
present section before going on to study the three-point function
in \S\ref{sec:three-point}.

After expanding $S$ according to its definition, it follows that
the third order term can be written
\begin{eqnarray}
	\fl\nonumber
	S_{3} = \frac{1}{2} \int dt \, d^{3}x \; e^{3\rho} \sqrt{\gamma} \Big( -
	\frac{1}{3} V_{0}''' \varphi^{3} -V_{0}'' \delta N \varphi^{2}
	- \delta N \dot{\varphi}^{2}-2\dot{\varphi} N^{i} \varphi_{,i}
	\\ \nonumber \qquad\qquad \mbox{}
	- \delta N e^{-2 \rho} \gamma^{i j} \varphi_{,i} \varphi_{,j}
	+ 6\dot{\rho}^{2} \delta N^{3} +4 \dot{\rho} \triangle \chi \delta N^{2}
	- \delta N \chi_{|ij} \chi^{|ij}
	\\ \qquad\qquad \mbox{}
	+ \delta N(\triangle \chi)^{2}
	+ 2 \dot{\phi} \delta N N^{i} \varphi_{,i} -\dot{\phi}^{2} \delta N^{3}
	+ 2\dot{\phi} \delta N^{2} \dot{\varphi}
	\Big) ,
	\label{eq:three-point-action}
\end{eqnarray}
where to avoid unnecessary clutter we have denoted the covariant
derivative on the unit three-sphere by a vertical bar, so that
$X_{|i} \equiv \nabla_i X$ and
$X^{|i} \equiv h^{ij} X_{|j}$.

From this expression we require the leading order term in slow-roll.
This will allow us to compute the expectation value
$\langle \varphi \varphi \varphi \rangle$ to the first non-trivial
order at horizon crossing, after which we must perform a gauge transformation
to determine $\langle \zeta \zeta \zeta \rangle$.
After horizon crossing we expect the correlation functions of
the curvature perturbation to be approximately conserved.

To identify the leading slow-roll terms, we must relate the derivatives
of the potential to the motion of the background field $\phi$.
For this purpose one can use the relations
\begin{equation}
	V_{0}''' = \Big( \frac{6 \dot{\rho}}{\dot{\phi}}
	-3 \frac{\ddot{\phi}}{\dot\phi^{2}} \Big)
	e^{-2\rho} + \Or(\epsilon^{3/2}) ,
	\quad \mbox{and} \quad
	V_{0}'' = -3 e^{-2 \rho} + \Or(\epsilon) .
	\label{eq:order-estimate}
\end{equation}
All other terms in Eq.~\eref{eq:three-point-action} are suppressed by
at least one power of $\dot{\phi}/\dot{\rho}$, so
the leading contribution to the action comes from $V_0'''$.
On flat spatial slices this term is usually negligible
\cite{Falk:1992sf,Seery:2008qj}, since it is proportional to
powers of $\dot{\phi}$ and $\ddot{\phi}$.
These terms are indeed present in Eq.~\eref{eq:order-estimate},
but are accompanied by a term of order $\dot{\rho}/\dot{\phi}$
whose source is the curvature term in the background scalar field
equation, Eq.~\eref{eq:friedmann}.
Accordingly, a significant three-point interaction can be present at early
times when the scalar field is behaving in a way quite different from the
flat slicing expectation.

We are assuming that a slow-roll hierarchy exists at the time of
evaluation, so it follows that the $\ddot{\phi}$ contribution can be
discarded and the leading contribution to the action can be written
\begin{equation}
	\label{action_dom}
	S_{3} = -\int dt \, d^{3}x \; \sqrt{\gamma} e^{\rho} \;
	\frac{\dot{\rho}\varphi^3}{\dot{\phi}} .
\end{equation}
This is of order $\dot{\rho}/\dot{\phi} \sim \epsilon^{-1/2}$ and is
a qualitatively new contribution which is not present in the
interactions among $\varphi$ quanta on $\mathbb{R}^3$ spatial slices
\cite{Maldacena:2002vr}. This term is suppressed by
two powers of the scale factor compared to the vacuum energy density
and therefore appears in
Eq.~\eref{eq:order-estimate} proportional to $e^{-2\rho}$, which implies
that at late times where $e^{\rho} \rightarrow \infty$ it no longer
contributes to the interactions. Thus, for large $k$ we can expect that the
effect of primordial curvature disappears and we recover the standard
flat space result.

In \S\ref{sec:three-point} we will determine the contribution
that this interaction makes to the $\langle \zeta \zeta \zeta \rangle$
expectation value. In fact,
we will see that it gives rise to a term which
behaves like $\epsilon^{-2}$, and which can therefore
dominate over those terms which are common to flat hypersurfaces
and $S^3$ hypersurfaces (to be described below),
which behave at leading order like $\epsilon^{-1}$.
Of course, as we have described, other terms in the slow-roll expansion of
$V_{0}'''$ will dominate if $e^{-2\rho}$ is small---that is,
if the curvature of the spatial hypersurfaces
is small. We can only expect the term in equation~\eref{action_dom} to dominate in the three point function if there has not been much
inflation prior to the mode leaving the horizon.

The next-to-leading order terms contribute proportional to
$\epsilon^{1/2}$. Their effect in the action can be written
\begin{equation}
	\fl
	\label{action_flat}
	S_{3} = \frac{1}{2} \int dt\, d^{3}x \; e^{3 \rho} \sqrt{\gamma} \Big(
	\frac{\ddot\phi}{\dot\phi^2} e^{-2 \rho} \varphi^3
	+ 3 e^{-2 \rho} \delta N \varphi^{2}
	- \delta N \dot\varphi^{2}
	- 2 \dot\varphi N^{i} \varphi_{| i}
	- \delta N e^{-2 \rho} \varphi^{| i} \varphi_{| i}\Big) .
\end{equation}
These terms give rise to contributions which are common to both $S^3$ and
$\mathbb{R}^3$, and are suppressed by a power of $\epsilon$ compared to the
leading curvature term~\eref{action_dom}. They are the same as the terms
obtained by Maldacena \cite{Maldacena:2002vr}.
Adding both these contributions together, integrating by parts,
and using the background equations for $\rho$ and $\phi$, we obtain 
\begin{equation}
	\label{total_action}
	\fl
	S_{3} = - \int dt\, d^{3}x \; \left\{ \sqrt{\gamma} \; 
	\Big(
	e^{\rho}\, \frac{\dot{\rho}}{\dot{\phi}}\, \varphi^{3}
	+ e^{5 \rho} \dot\phi \dot\varphi^2 (\triangle +3)^{-1} \dot\varphi
	\Big)
	+ \frac{\delta L_{2}}{\delta \varphi} f(\varphi)
	\right\}
\end{equation}
where $f(\varphi)$ is an auxiliary function defined by
\begin{equation}
	\fl
	\label{field_redefinition}
	f(\varphi) \equiv
	\frac{\dot \phi}{8 \dot \rho} \varphi^2
	- \frac{3 \dot \phi}{8 \dot \rho} (\triangle +3)^{-1} (\varphi^2)
	- \frac{\dot \phi}{4 \dot \rho} (\triangle +3)^{-1}
	\left\{ \varphi (\triangle +3) \varphi \right\} + \cdots
\end{equation}
and `$\cdots$' represents terms which vanish at late times.
The term involving $\delta L_2/\delta \varphi$ is proportional to the
leading-order equation of motion and therefore vanishes when we take
the interaction picture field to be on-shell. However, it cannot simply
be discarded; it records the contribution of boundary terms which
were generated after integrating by parts and which we have not written
explicitly \cite{Seery:2006tq}. The correct procedure is to make a field
redefinition to remove these terms, by introducing a shifted field
$\varphi_c$, defined by
\begin{equation}
	\label{field_redefinition2}
	\varphi = \varphi_{c} + f(\varphi_{c}) .
	\end{equation}
This removes the nuisance terms proportional to the equation of motion
\emph{and} the boundary terms, giving a simplified action
\begin{equation}
	\label{action_refined}
	S_{3} = -\int dt \, d^{3}x \;
	\sqrt{\gamma} \;
	\left(  
	e^{\rho}\, \frac{\dot{\rho}}{\dot{\phi}}\, \varphi_{c}^{3}
	+ e^{5 \rho} \dot\phi \dot\varphi_{c}^2 (\triangle +3)^{-1}
	\dot\varphi_{c} 
	\right) .
\end{equation}
We may now compute the simpler correlation functions of
$\varphi_c$, and rewrite the result in terms of the interesting field
$\varphi$ by using Eq.~\eref{field_redefinition2}.

\section{The three-point function}
\label{sec:three-point}
\subsection{The $\varphi$ correlation function}

The final step is to calculate the three-point function in the
comoving gauge~\eref{comoving_gauge}. As we have described,
the comoving curvature perturbation $\zeta$ is conserved after horizon
exit in the absence of non-adiabatic pressure.
Once a given scale
falls back inside the horizon, $\zeta$ can be used to seed the subsequent
calculation of temperature and density fluctuations in the coupled
baryon--photon plasma.

To obtain the correlation functions of $\zeta$ entails a number of steps.
The first
requires that we obtain the contribution from the reduced
third order action~\eref{action_refined},
and rewrite it in terms of the correlation functions of the original
field $\varphi$ using the field redefinition~\eref{field_redefinition}.
Once this has been done it is necessary to determine how the
correlation functions of
$\zeta$ are related to those of $\varphi$. Fortunately, there is a simple
prescription (the so-called ``$\delta N$ formalism'')
which is valid on large scales
\cite{Starobinsky:1986fx,Sasaki:1995aw,Lyth:2004gb,Lyth:2005fi}.
The result is that $\zeta(t,\vec{x})$ evaluated at any time $t$ later than the
horizon-crossing time $t_\ast$ can be written
\begin{equation}
	\zeta(t,\vec{x}) \equiv \delta N(t,\vec{x}) =
	\sum_{n=1}^{\infty} \frac{1}{n!}
		\left\{ \frac{\partial^n}{\partial \phi_\ast^n} N(t,t_\ast) \right\}
	[\varphi_\ast(t_\ast,\vec{x})]^n ,
	\label{eq:delta-N}
\end{equation}
where $t_\ast$ is taken to be a spatial slice on which the curvature perturbation $\zeta$ vanishes, $t$ labels a slice
of uniform energy density, and $N$ is the number of e-folds between these two slices. Note that this formula applies only in
coordinate space and must be treated accordingly when transforming to
harmonic modes.
Eq.~\eref{eq:delta-N} can also be understood in terms of a
further field redefinition which
changes the action from the uniform curvature to comoving gauge
\cite{Maldacena:2002vr}.
The whole computation is performed with the interaction picture fields using either~\eref{mode_approx} or~\eref{mode} according to whether $k$ is large
or small.

Let us first determine the contribution from the first term in the reduced
third-order action, Eq.~\eref{action_refined}.
Provided we are only interested in tree-level amplitudes, the interaction
Hamiltonian is simply given by minus the interaction term in the
Lagrangian \cite{Seery:2007we,Dimastrogiovanni:2008af},
so that $H_{\intr \; 3} = - L_{\intr \; 3}$.
It follows that the contribution from the first term
in~\eref{action_refined}, for large $k$, is
\begin{eqnarray}
	\fl\nonumber
	\langle
	\varphi_{\vec{k}_{1}}\varphi_{\vec{k}_{2}}\varphi_{\vec{k}_{3}}
	\rangle
	\supseteq
	 - \imath \int_{\imath \infty}^{\pi/2 - \varepsilon} d \eta \;
	\langle [\varphi_{\vec{k}_{1}} ( \pi /2 )
		\varphi_{\vec{k}_{2}} ( \pi / 2)
		\varphi_{\vec{k}_{3}} ( \pi / 2) , H_{\intr \;3}(\eta)
	] \rangle \\
	= \cdots - \frac{6 \dot{\rho}_{*}^{5} \lambda^{4}}
		{8 k_{1}^{2} k_{2}^{2} k_{3}^{2} \dot{\phi}_{*}}
	e^{{-\imath \pi k_t / 2}}
	\int d^{3}x \; \sqrt{\gamma}
	Q_{\vec{k}_{1}} Q_{\vec{k}_{2}} Q_{\vec{k}_{3}} J
	+ \mbox{c.c.} ,
	\label{eq:primitive-three-a}
\end{eqnarray}
where ``c.c.'' denotes the complex conjugate of the preceding expression
and $k_{t} \equiv k_{1}+k_{2}+k_{3}$.
The function $J$ is defined by
\begin{equation}
	J = \int_{\imath \infty }^{\pi/2-\varepsilon} d\eta \;
	\frac{\sin \eta}{\cos^2 \eta}
	\Big(\cos \eta - \frac{\imath}{k_1} \Big)
	\Big(\cos \eta - \frac{\imath}{k_2} \Big)
	\Big(\cos \eta - \frac{\imath}{k_3} \Big)
	e^{\imath k_t \eta} .
	\label{eq:primitive-three-b}
\end{equation}
In Eqs.~\eref{eq:primitive-three-a}--\eref{eq:primitive-three-b}
we have carried the integration over conformal time to within a small
parameter $\varepsilon$ (not to be confused with the slow-roll
parameter $\epsilon$) of future infinity.
We can expect that this will be a good approximation because
the integral receives its largest contribution from around the time of
horizon exit and thereafter does not evolve appreciably, so there is
little error in continuing the integration into the infinite future.
Inside the horizon the integral is taken over a contour which turns
a right-angle at $\eta = 0$ and approaches infinity along the positive
imaginary axis, corresponding to evaluation of the interaction in the
Hartle--Hawking state. This is equivalent to the interacting vacuum of
the full theory.

Eqs.~\eref{eq:primitive-three-a}--\eref{eq:primitive-three-b}
give rise to a term in the $\varphi$ correlation function which takes the
form
\begin{equation}
	\fl
	\langle
	\varphi_{\vec{k}_{1}}\varphi_{\vec{k}_{2}}\varphi_{\vec{k}_{3}}
	\rangle
	\supseteq
	- \frac{3 \dot{\rho}_{*}^{5}}
	{2 \dot{\phi}_{*} k_{1}^{3} k_{2}^{3} k_{3}^{3}}
	\int d^3 x \; \sqrt{\gamma} 
	Q_{\vec{k}_{1}} Q_{\vec{k}_{2}} Q_{\vec{k}_{3}}
	\Big( -2 \frac{k_1 k_2 k_3}{k_t^2}
		+ k_t
		- \frac{1}{k_t^2} \sum_{i \ne j} k_i k_j^2
	\Big)
	\label{eq:primitive-three-c}
\end{equation}
In order to obtain the contribution this makes to the $\zeta$ correlation
funtion we will shortly see that the $\delta N$ formalism tells us we must
multiply by $(-\dot\rho_* / \dot\phi_*)^3$. The result will be proportional
to $\epsilon^{-2} k^{-8} \dot{\rho}_{*}^4$, whereas the usual terms from the
calculation with flat constant time hypersurfaces are proportional to
$\epsilon^{-1} k^{-6} \dot{\rho}_{*}^4$. So we expect the terms given in
Eq.~\eref{eq:primitive-three-c} to dominate for small $k$.

On the other hand, some of the usual terms in the three point function are
obtained by calculating the contribution from the second term in Eq.~\eref{action_refined}. These terms are given by
\begin{equation}
	\fl
	\langle
	\varphi_{\vec{k}_{1}}\varphi_{\vec{k}_{2}}\varphi_{\vec{k}_{3}}
	\rangle
	\supseteq
	- \frac{\dot{\phi}_{*} \dot{\rho}_{*}^3}
	{2 k_{1}^{3} k_{2}^{3} k_{3}^{3}} 
	\int d^3 x \; \sqrt{\gamma}
	Q_{\vec{k}_{1}} Q_{\vec{k}_{2}} Q_{\vec{k}_{3}}
	\sum_{i>j} \frac{k_i^2 k_j^2}{k_t} .
\end{equation}
Adding on the terms from the field redefinition (see section~\eref{section:field_redefinition_terms}), given by
Eqs.~\eref{field_redefinition} and~\eref{field_redefinition2}, and taking the dominant terms for large $k$, we
obtain 
\begin{equation}
	\fl
	\langle
	\varphi_{\vec{k}_{1}}\varphi_{\vec{k}_{2}}\varphi_{\vec{k}_{3}}
	\rangle
	\supseteq
	- \frac{\dot{\phi}_{*} \dot{\rho}_{*}^{3}}
	{8 k_{1}^{3} k_{2}^{3} k_{3}^{3}}
	\int d^3 x \; \sqrt{\gamma} 
	Q_{\vec{k}_{1}} Q_{\vec{k}_{2}} Q_{\vec{k}_{3}}
	\Big( \frac{4}{k_t} \sum_{i>j} k_i^2 k_j^2
		- \frac{1}{2} \sum_{i} k_i^3
		+ \frac{1}{2} \sum_{i \ne j}k_i k_j^2
	\Big)
\end{equation}

\subsection{Terms due to the field redefinition}
\label{section:field_redefinition_terms}
The field redefinition given by Eqs.~\eref{field_redefinition}--\eref{field_redefinition2} contributes a number of terms to the three point function. These are calculated with the aid of a convolution (see Eq.~\eref{eqn:convolution}) and then contraction of the fields in such a way that only connected diagrams are produced. The first term of Eq.~\eref{field_redefinition} results in a calculation much like that which will be done in section \S\ref{sec:deltaN}. The second and third terms in Eq.~\eref{field_redefinition} are given by a marginally more complicated calculation where one must integrate by parts to throw the $(\Delta +3)^{-1}$ onto the $Q_{\vec{k}}$ in the convolution, and then contract the fields. The sum total of the redefinition, Eq.~\eref{field_redefinition}, is then
\begin{eqnarray}
	\nonumber
	\fl
	\langle
	\varphi_{\vec{k}_{1}}\varphi_{\vec{k}_{2}}\varphi_{\vec{k}_{3}}
	\rangle
	-&
	\langle
	\varphi_{c\,\vec{k}_{1}}\varphi_{c\,\vec{k}_{2}}\varphi_{c\,\vec{k}_{3}}
	\rangle
	=\\
	&
	\frac{\dot\phi_* \dot\rho_*^3 \lambda^4}{16 \prod_l k_l^3}
	\int d^3 x \; \sqrt{\gamma} 
	Q_{\vec{k}_{1}} Q_{\vec{k}_{2}} Q_{\vec{k}_{3}}
	\left(
	\sum_i k_i^3 + 3 \sum_i k_i - \sum_{i\ne j} k_i^2 k_j
	\right)
\end{eqnarray}

\subsection{The $\delta N$ formalism}
\label{sec:deltaN}
Finally, one uses the $\delta N$ prescription to obtain the full
$\zeta$ correlation function. Multiplying three copies of
Eq.~\eref{eq:delta-N}, taking correlations using Wick's theorem
and truncating to tree-level terms which do not involve unconstrained
momentum integrations, we find
\begin{equation}
	\fl
	\langle
		\zeta_{\vec{k}_{1}}\zeta_{\vec{k}_{2}}\zeta_{\vec{k}_{3}}
	\rangle
	=
	\left(\frac{\delta N}{\delta \phi_\ast}\right)^3
	\langle
		\varphi_{\vec{k}_{1}}\varphi_{\vec{k}_{2}}\varphi_{\vec{k}_{3}}
	\rangle
	+ \frac{1}{2}
	\left( \frac{\delta N}{\delta \phi_\ast} \right)^2
	\frac{\delta^2 N}{\delta\phi_\ast^2}
	\left[
		\langle\varphi_{\vec{k}_{1}}\varphi_{\vec{k}_{2}}
		(\varphi\star\varphi)_{\vec{k}_{3}}\rangle
		+ \mbox{cyclic}
	\right],
\end{equation}
where $\star$ denotes a `convolution,'
\begin{equation}
\label{eqn:convolution}
(\varphi\star\varphi)_{\vec{k}}(t)\equiv \int d^3 x \sqrt{\gamma} Q_{\vec{k}}^{*}(\vec{x}) \varphi^2 (t,\vec{x}),
\end{equation}
 ``cyclic'' indicates that all cyclic 
permutations of $\{ 1, 2, 3 \}$ should be included in the sum, 
and $N$ measures by how many e-folds the mode in question is outside 
the horizon (see description below Eq.~\eref{eq:delta-N}). 
Since we are working with a single-field model of inflation, 
the derivatives of $N$ can be evaluated directly, yielding 
\begin{equation}
	\frac{\delta N}{\delta \phi_\ast} =
	- \frac{\dot{\rho}_\ast}{\dot{\phi}_\ast} ,
	\quad \mbox{and} \quad
	\frac{\delta^2 N}{\delta \phi_\ast^2} =
	\frac{\dot{\rho}_\ast \ddot{\phi}_\ast}{\dot{\phi}_\ast^3}
	+ \frac{1}{2}
	- \frac{1}{\dot{\phi}_\ast^2 e^{2 \rho}_\ast} .
\end{equation}

\subsection{The $\zeta$ correlation function}
Collecting all these terms, the final result for the three point function evaluated on a late time slice is
\begin{eqnarray}
	\fl\nonumber
	\langle
	\zeta_{\vec{k}_{1}}\zeta_{\vec{k}_{2}}\zeta_{\vec{k}_{3}}
	\rangle =
	\frac{\dot{\rho}_{*}^{6}}
	{8 \dot{\phi}_{*}^{2} k_{1}^{3} k_{2}^{3} k_{3}^{3}}
	\int d^3 x \; \sqrt{\gamma}
	Q_{\vec{k}_{1}} Q_{\vec{k}_{2}} Q_{\vec{k}_{3}} \\ \nonumber \qquad
	\mbox{} \times \Bigg[
		\frac{4}{k_t} \sum_{i>j} k_i^2 k_j^2
		+ \frac{1}{2} \sum_{i} k_i^3
		+ \frac{1}{2} \sum_{i \ne j} k_i k_j^2
		+ \frac{2\dot{\rho}_* \ddot{\phi}_*}{\dot{\phi}_*^3}
			\sum_{i} k_i^3
	\\ \qquad\qquad \mbox{}
	+ \frac{2 \dot{\rho}_*^2}{\dot{\phi}_*^2}
	\Big( -12 \frac{k_1 k_2 k_3}{k_t^2}
		+ 6 k_t
		- \frac{6}{k_t^2} \sum_{i \ne j} k_i k_j^2
		-\frac{1}{\dot\rho_{*}^{2} e^{2 \rho_*}} \sum_i k_i^3
	\Big)
	\Bigg]
	\label{final_correlator}
\end{eqnarray}
Of course, this uses the large $k$ approximation for $\varphi^{\cl}$ and so
it can only be regarded as approximate.

The leading term in this expression scales with momentum like $[k^{-6}]$.
This is the term computed by Maldacena, and is dominant for sufficiently
large $k$, that is, on small scales.
The correction terms we have computed scale with momentum like
$[k^{-8}]$. These terms can become dominant on larger scales.
On sufficiently large scales, of course, one must remember that
Eq.~\eref{final_correlator} will be accompanied by other corrections
that scale even faster as powers of momentum, like $[k^{-2n}]$ for
any integer $n \geq 3$, and that these corrections will eventually
overwhelm the ones we have computed. In this regime, for accuracy, one 
should use the full expression for the modes~\eref{mode} and perform a 
numerical assessment of the three point function.

\subsection{An estimate of $\fnl$}
\label{sec:projection}
We are finally in a position to estimate the magnitude of the observable
non-linearity parameter $\fnl$ which is produced by the sensitivity to
curvature in Eq.~\eref{final_correlator}. We define the sign of $\fnl$
according to the WMAP convention, where the bispectrum is parameterized
in harmonic space via
\begin{equation}
	B(k_1,k_2,k_3) =
	\frac{6}{5} \fnl \left\{
		P(k_1) P(k_2) + P(k_1) P(k_3) + P(k_2) P(k_3)
	\right\} .
	\label{eq:fnl-def}
\end{equation}
One must be careful in interpreting this formula for modes on $S_3$, because
when we work with the harmonics $Q_{\vec{k}}$, the $k_i$ no longer have the
same meaning as in flat space. The correct way to compare the magnitude of
$\fnl$ between models with different spatial geometries is to project
onto the sky and compare the non-linearity parameter in the angular
expectation values of $\zeta$.

To obtain the correct projection, we begin by writing
the three-$\zeta$ correlator in terms of a function $\xi(k_1,k_2,k_3)$
defined by
\begin{equation}
	\langle
	\zeta_{\vec{k}_{1}}\zeta_{\vec{k}_{2}}\zeta_{\vec{k}_{3}}
	\rangle
	=
	\int d^3 x \; \sqrt{\gamma}
	Q_{\vec{k}_{1}} Q_{\vec{k}_{2}} Q_{\vec{k}_{3}} \xi(k_1, k_2, k_3) .
\end{equation}
To obtain the angular correlation function of the $\zeta$s,
we follow the route outlined in
\S\ref{sec:sky}: the correlator in harmonic space is transformed back to
coordinate space, and evaluated at the radial distance corresponding to
last scattering. In terms of the polar coordinate on $S^3$, this is
given by $\chi = \chi_L$. One finds
\begin{eqnarray}
	\fl\nonumber
	\langle
	\zeta_{l_1 m_1} \zeta_{l_2 m_2} \zeta_{l_3 m_3} \rangle_{\chi_L}
	= C_{l_1\;\;\, l_2\;\;\, l_3}^{m_1 m_2 m_3}
	\sum_{k_i \geq (l_i + 1)}
	\Pi_{k_1}^{l_1} (\chi_L)
	\Pi_{k_2}^{l_2} (\chi_L)
	\Pi_{k_3}^{l_3} (\chi_L) \\	
	\times	
	\int_0^\pi \sin^2 \chi \; d \chi \;
	\Pi_{k_1}^{l_1} (\chi)
	\Pi_{k_2}^{l_2} (\chi)
	\Pi_{k_3}^{l_3} (\chi)
	\xi(k_1, k_2, k_3) 
\end{eqnarray}
where
\begin{equation}
	C_{l_1\;\;\, l_2\;\;\, l_3}^{m_1 m_2 m_3} =
	\int d \Omega^2(\hat{\vec{x}}) \;
	Y_{l_1}^{m_1} (\hat{\vec{x}})
	Y_{l_2}^{m_2} (\hat{\vec{x}})
	Y_{l_3}^{m_3} (\hat{\vec{x}}) ,
\end{equation}
and $d \Omega^2(\hat{\vec{x}})$ is an element of solid angle on $S^2$ in the direction 
of the unit vector $\hat{\vec{x}}$. The integration over $\chi$ is 
symmetric or antisymmetric about $\chi = \pi/2$, with the two regions 
interfering constructively for $k_t$ odd and destructively for $k_t$ even. 
Moreover, the $\Pi_k^l(\chi)$ under the integral can be replaced by 
spherical Bessel functions for the range of $\chi$, $k$ and $l$ of interest. 
It follows that
\begin{eqnarray}
	\fl\nonumber
	\langle
	\zeta_{l_1 m_1} \zeta_{l_2 m_2} \zeta_{l_3 m_3} \rangle_{\chi_L}
	\approx
	C_{l_1\;\;\, l_2\;\;\, l_3}^{m_1 m_2 m_3} 
	\Big( \frac{2}{\pi} \Big)^3
	\int_0^{\pi / 2} \sin^2 \chi \; d \chi
	\\ \nonumber
	\mbox{} \times
	\Bigg\{
		\sum_{k_1 \geq (l_1 + 1)} k_1^2 j_{l_1} (k_1 \chi)
			j_{l_1} (k_1 \chi_L)
	\Bigg\}
	\Bigg\{ 1 \leftrightarrow 2 \Bigg\}
	\Bigg\{ 1 \leftrightarrow 3 \Bigg\}
	\\
	\mbox{} \times
	[1-(-1)^{k_t}] \xi(k_1,k_2,k_3)
\end{eqnarray}
The influence of the $[1-(-1)^{k_t}]$ term averages out to $1$ as we sum 
over, say, $k_3$ and the summation over the $k_i$ may be replaced by
integrals to give
\begin{eqnarray}
	\fl\nonumber
	\langle
	\zeta_{l_1 m_1} \zeta_{l_2 m_2} \zeta_{l_3 m_3} \rangle_{\chi_L} 
	\approx 
	C_{l_1\;\;\, l_2\;\;\, l_3}^{m_1 m_2 m_3}
	\Big( \frac{2}{\pi} \Big)^3
	\int_0^{\pi / 2} \sin^2 \chi \; d \chi
	\\ \label{eq:angular-bispectrum-intermediate}
	\mbox{} \times
	\Bigg\{
		\int_{(l_1 + 1)}^\infty d k_1 \; k_1^2 j_{l_1} (k_1 \chi)
		j_{l_1} (k_1 \chi_L)
	\Bigg\}
	\Bigg\{ 1 \leftrightarrow 2 \Bigg\}
	\Bigg\{ 1 \leftrightarrow 3 \Bigg\}
	\xi(k_1,k_2,k_3) 
\end{eqnarray}
First consider the part of $\xi$ which scales with momentum like
$[k^{-6}]$, coming from those terms in Eq.~\eref{final_correlator}
which were computed by Maldacena and are present in the flat space
expectation value. In order to carry out the integrations in
Eq.~\eref{eq:angular-bispectrum-intermediate} explicitly
it is most convenient if the terms involving each $k_i$ in
$\xi$ factorize, so that each $k_i$ integral can be evaluated independently.
The necessary components of $\xi$ are
\begin{equation}
	\label{xi_large_k}
	\fl
	\xi(k_1,k_2,k_3)
	=
	\frac{\dot{\rho}_{*}^{6}}
	{8 \dot{\phi}_{*}^{2} k_{1}^{3} k_{2}^{3} k_{3}^{3}}
	\Big(
		\frac{4}{k_t} \sum_{i>j} k_i^2 k_j^2
		+ \frac{1}{2} \sum_{i} k_i^3
		+ \frac{1}{2} \sum_{i \ne j} k_i k_j^2
		+ \frac{2\dot{\rho}_* \ddot{\phi}_*}{\dot{\phi}_*^3}
			\sum_{i} k_i^3
	\Big)+\ldots
\end{equation}
There is a small variation with $k$ owing to the variation of
$\dot{\rho}_\ast$, $\dot{\phi}_\ast$ and $\ddot{\phi}_\ast$ with the
epoch of horizon exit. Ignoring this slow variation, however,
the only term which does not factorize involves $1/k_t$.
A similar obstruction was encountered by Smith \& Zaldarriaga
\cite{Smith:2006ud}, who found a factorizable form by
introducing a Schwinger parameter,
\begin{equation}
	\frac{1}{k_t} \equiv \int_0^\infty d t \; e^{-t (k_1 + k_2 + k_3)} .
\end{equation}

To understand which regions make significant contributions to the
integral in
Eq.~\eref{eq:angular-bispectrum-intermediate}
one can account for the influence of the Bessel functions
using a stationary phase method.
In practice, one employs a WKB approximation
to write these functions as an amplitude multiplied by a term with 
rapidly varying phase.
An appropriate WKB solution can be constructed from the defining
equation of the spherical Bessel functions,
\begin{equation}
	\frac{d^2}{d z^2}
	\left[ z j_l (z) \right]
	=\Big({l(l+1)\over z^2} -1 \Big)
	\left[ z j_l(z) \right] .
\end{equation}

The derivative of the WKB phase satisfies
\begin{equation}
	\frac{d}{d k}(\textrm{WKB phase})
	=
	\frac{d}{d k} \int_{\sqrt{l(l+1)}}^{k\chi} \sqrt{1-{l(l+1)\over z^2}}
	\; d z
	=
	\chi \sqrt{1-{l(l+1)\over k^2 \chi^2}}.
\end{equation}
First consider contributions to
the $\chi$ integral in Eq.~\eref{eq:angular-bispectrum-intermediate}
from the region where $\chi \gg \chi_L$.
There are two cases, depending whether $k^2$ is larger or smaller than
$l(l+1)/\chi^2$.
For very large $k$, the
stationary phase approximation implies that
there is essentially no contribution to the integral.
We are therefore left with the region $k^2 \le l(l+1)/\chi^2$.
Since $\chi \gg \chi_L$ it follows that $k^2 \chi_L^2 \ll l(l+1)$, and
therefore $j_l (k \chi_L)$ is becoming exponentially suppressed.
It follows that there is also a negligible contribution in this case.
The only significant contribution comes from the region where
$\chi \lesssim \chi_L$.
In this region we can approximate
$\sin \chi \approx \chi$ because $\chi_L \ll 1$; the error this induces
in the region $\chi > \chi_L$ is immaterial because the integrand is
exponentially suppressed there.
Moreover, 
in view of what has been said about the region where $\chi \gg \chi_L$,
there is very little
error in extending the range of $\chi$ integration to infinity.
Accordingly, we have
\begin{eqnarray}
	\fl\nonumber
	\langle \zeta_{l_1 m_1} \zeta_{l_2 m_2} \zeta_{l_3 m_3} \rangle_{\chi_L} 
	\approx 
	C_{l_1\;\;\, l_2\;\;\, l_3}^{m_1 m_2 m_3}
	\Big( \frac{2}{\pi} \Big)^3
	\int_0^{\infty} \chi^2 \; d \chi \\
	\times	
	\Bigg\{
		\int_{(l_1 +1)}^{\infty} d k_1 \; k_1^2 j_{l_1} (k_1 \chi)
		j_{l_1} (k_1 \chi_L)
	\Bigg\}
	\Bigg\{ 1 \leftrightarrow 2 \Bigg\}
	\Bigg\{ 1 \leftrightarrow 3 \Bigg\}
	\xi(k_1,k_2,k_3) .
\end{eqnarray}
We are assuming that $\xi$ is factorizable, but if it is not
a similar discussion applies after the introduction of Schwinger parameters.

Let us relate this expression to the flat space limit, where $\chi_L \ll 1$. 
There is only an exponentially suppressed contribution to each $k$ 
integral from the region $k_i \in [0, l_i + 1]$, because if $\chi_L \ll 1$ 
then it must also be true that $k_i^2 \chi_L^2 \ll l_i (l_i + 1)$ 
for each $i$, 
and it follows that the Bessel function $j_{l_i}(k_i \chi_L)$ is undergoing 
exponential suppression in this region. We therefore incur essentially 
no penalty in extending the lower limit of integration to $0$, giving 
\begin{eqnarray}
	\fl\nonumber
	\langle \zeta_{l_1 m_1} \zeta_{l_2 m_2} \zeta_{l_3 m_3} \rangle_{\chi_L}
	\approx 
	C_{l_1\;\;\, l_2\;\;\, l_3}^{m_1 m_2 m_3}
	\Big( \frac{2}{\pi} \Big)^3
	\int_0^{\infty} \chi^2 \; d \chi \\
	\mbox{} \times
	\Bigg\{
		\int_{0}^{\infty} d k_1 \; k_1^2 j_{l_1} (k_1 \chi)
		j_{l_1} (k_1 \chi_L)
	\Bigg\}
	\Bigg\{ 1 \leftrightarrow 2 \Bigg\}
	\Bigg\{ 1 \leftrightarrow 3 \Bigg\}
	\xi(k_1,k_2,k_3) 
\end{eqnarray}
This is the standard formula for the angular bispectrum of the
curvature perturbation with flat spatial slices. It follows that in
the approximate flat space limit, we can compare $\xi$ between the
cases of $\mathbb{R}^3$ and $S^3$ spatial slices to find an estimate
of the comparative magnitude of $\fnl$. One finds
\begin{eqnarray}
	\nonumber\fl
	\fnl
	= - \frac{5}{12 \sum_l k_l^3}
	\Big[
		\frac{\dot\phi_*^2}{\dot\rho_*^2}
		\Big( \frac{4}{k_t} \sum_{i>j} k_i^2 k_j^2
			+ \frac{1}{2} \sum_i k_i^3
			+ \frac{1}{2} \sum_{i\ne j} k_i k_j^2
			+ \frac{2\dot\rho_* \ddot \phi_*}{\dot\phi_*^3}
				\sum_i k_i^3
		\Big) \\ \qquad
	\mbox{} +
	2 \Big( -12 \frac{k_1 k_2 k_3}{k_t^2}
		+ 6 k_t
		- \frac{6}{k_t^2} \sum_{i \ne j} k_i k_j^2
		-\frac{1}{\dot\rho_*^2 e^{2 \rho_*}}\sum_i k_i^3
	\Big)
	\Big] .
	\label{fnl}
\end{eqnarray}
The correction terms can be larger than the flat space result when
$k \lesssim \epsilon^{-1/2}$.

\section{Conclusions}
\label{sec:conclusions}

We work in the scenario where the Universe is a slowly rolling
but positively curved spacetime formed from a background with
constant density spatial slices which are copies of $S^3$, as opposed to $\mathbb{R}^3$.
This possibility is compatible with observation if
$|\Omega_k| \lesssim 10^{-2}$--$10^{-3}$, depending whether the universe
is taken to be positively or negatively curved.
Indeed, interpreting present observational limits literally,
the diameter of the CMB $d_{\ell=2}$ satisfies approximately
$d_{\ell=2}\approx 0.46$ (in units where the radius of curvature
is unity), so that the $S^2$ corresponding to
the surface of last scattering spans about a twelth of the 
circumference of a great circle on the $S^3$ of constant time. On such a last 
scattering wall one can expect that the fluctuation eigenmodes on $S^3$ of 
$k\gtrsim 8$ will contribute to the CMB.
Modes of long wavelength, corresponding to small $k$, will pass outside
the horizon early and therefore will be sensitive the presence of curvature.
The question is to what degree a realistic observable such as the power
spectrum, or $\fnl$, can constrain the appearance of primordial curvature.
We have shown that such modes, of small $k$, will not contribute
significantly
to the spectrum, but may contribute to $\fnl$.

We perform the appropriate selection of the vacuum for our slowly rolling de 
Sitter type spacetime, corresponding to the Hartle--Hawking state,
and demonstrate how one should execute the calculation 
of both the two and three point functions on this manifold. Along the way 
we find exact expressions for the scalar fluctuation modes in global de 
Sitter coordinates, and since these are rather unwieldy we explore appropriate  
approximations. For a generic potential $V$ one finds that the contributions 
to the action due to curvature at leading order in slow roll are remarkably 
simple. Further, we show how the 
$S^3$ slicing can be reliably compared with the conventional flat-slicing
calculation on an $S^2$ surface corresponding to last scattering.
In this way we can establish continuity with the $\mathbb{R}^3$ calculation
as one approaches the flat-space limit.

We estimate the contribution of small-$k$ harmonics to the bispectrum.
To compare with observations, it is most convenient
to state the result in terms of $f_{NL}$, given in Eq.~\eref{fnl}.
As written, this expression should be understood to be valid
in the approximately equilateral case where all $k_i$ have
an approximately equal magnitude. However, we believe our calculation
could be generalized, using methods similar to those of
Maldacena \cite{Maldacena:2002vr}, to accommodate the squeezed limit.

How large an $\fnl$ can be obtained? The $k=1$ mode is the homogeneous
background, and $k=2$ is pure gauge. Therefore the interesting
contributions to microwave background fluctuations must come from
modes where $k \ge 3$.
To estimate an upper limit,
we specialize to the equilateral limit where $k_i = k$
for all $i$. After collecting terms in Eq.~\eref{fnl}, this yields
\begin{equation}
	\fnl = -\frac{5}{36} (23 \epsilon_* - 6 \eta_*) - \frac{145}{54 k^2} \, .
\end{equation}
The first term, proportional to the slow-roll parameters $\epsilon$ and
$\eta$, is precisely the flat-space contribution first derived by
Maldacena. The second term is new, and accounts for the leading effect of
curvature. It is irrelevant in the limit $k \rightarrow \infty$, but can
be large for small $k$.
Since the signal is maximized by choosing $k$
as small as possible, we can obtain an approximate upper bound by
setting $k = 3$, leading to $\fnl \sim 0.3$.
It follows that the effect of curvature is marginally below the expected
detection threshold for the \emph{Planck} satellite, usually supposed
to be of order $\fnl \sim 5$
\cite{Komatsu:2001rj}, but might lie on the limit of what is
practicable with futuristic technology such as
a high-redshift 21cm survey
\cite{Cooray:2006km,Pillepich:2006fj}.
In the latter case, however, one would need
some way to distinguish this small signal from the ubiquitous
non-linearities of gravity itself \cite{Boubekeur:2009uk}.

\ack

TC acknowledges support from EPSRC. DS is supported by STFC.

\section*{References}

\providecommand{\href}[2]{#2}\begingroup\raggedright\endgroup

\end{document}